\documentclass[12pt,fleqn]{article}
\usepackage[DIV12]{typearea}
\usepackage{amsmath,amsfonts,amssymb}
\usepackage{graphicx}
\usepackage{cite}
\usepackage{mathrsfs}
\usepackage{bbm}
\usepackage{pifont}
\usepackage{braket}
\usepackage{units}
\usepackage[small,loose]{subfigure}  % subfigures with (a), (b) etc., ... and own caption
\usepackage[dvipsnames]{xcolor}
\usepackage{mathtools}
\usepackage{tikz}
\usepackage[bookmarks=false]{hyperref}%\usepackage{hyperref}
\usepackage{enumerate}
\usepackage[normalem]{ulem}

\usepackage{pdfcomment}

\newcommand{\Apref}[1]{Appendix~\ref{#1}}

\newcommand{\Tabref}[1]{Table~\ref{#1}}

\newcommand{\eVdist}{\kern-0.06em}
\DeclareMathOperator{\tr}{tr}
\newcommand{\I}{\mathrm{i}}

\newcommand{\SU}[1]{\ensuremath{\mathrm{SU}(#1)}}
\newcommand{\U}[1]{\ensuremath{\mathrm{U}(#1)}}
\newcommand{\Z}[1]{\ensuremath{\mathbbm{Z}_{#1}}} % z_N ->\Z{N}

\newcommand{\Hu}{\ensuremath{H_u}}
\newcommand{\Hd}{\ensuremath{H_d}}

\newcommand{\singlet}{\ensuremath{N}}

\newcommand{\dilaton}{\ensuremath{S}}

\allowdisplaybreaks[1]

\numberwithin{equation}{section}
\numberwithin{table}{section}

\def\mytitle{Singlet extensions of the MSSM with $\Z4^R$ symmetry}

\title{\boldmath\mytitle\unboldmath}

\begin{document}

\begin{titlepage}

\begin{flushright}
TUM--HEP 984/15\\
LMU--ASC 09/15\\
FLAVOUR--EU 92
\end{flushright}

\vspace*{1.0cm}

\begin{center}
\typeout{\thefootnote}
\renewcommand{\thefootnote}{$\star$} 
{\Large\bf
\boldmath\mytitle\unboldmath
}
\renewcommand{\thefootnote}{\arabic{footnote}} 

\vspace{1cm}

\textbf{
Michael Ratz\footnote[1]{Email: \texttt{michael.ratz@tum.de}}{}$^a$ 
and
Patrick K.S.\ Vaudrevange\footnote[2]{Email: \texttt{patrick.vaudrevange@tum.de}}{}$^{b, c}$
}
\\[5mm]
\textit{\small
{}$^a$ Physik--Department T30, Technische Universit\"at M\"unchen, \\
~~James--Franck--Stra\ss e~1, 85748 Garching, Germany
}
\\[5mm]
\textit{$^b$\small
~Excellence Cluster Universe, \\
Boltzmannstra\ss e 2, 85748 Garching, Germany
}
\\[5mm]
\textit{$^c$\small
~Arnold Sommerfeld Center for Theoretical Physics,
Ludwig--Maximilians--Universit\"at M\"unchen, Theresienstra\ss e 37, 80333 M\"unchen, Germany
}

\end{center}

\vspace{1cm}

\begin{abstract}
We discuss singlet extensions of the MSSM with $\Z4^R$ symmetry.  We show that
holomorphic zeros can avoid a potentially large coefficient of the term linear
in the singlet. The emerging model has both an effective $\mu$ term and a
supersymmetric mass term for the singlet $\mu_{\singlet}$ which are controlled
by the gravitino mass. The $\mu$ term turns out to be suppressed against
$\mu_{\singlet}$ by about one or two orders of magnitude. We argue that this
class of models might  provide us with a solution to the little hierarchy
problem of the MSSM.
\end{abstract}

\end{titlepage}

\section{Purpose of this note}

The $\Z4^R$ symmetry \cite{Babu:2002tx,Lee:2010gv} provides us with compelling
solutions of the $\mu$ and proton decay problems of the minimal supersymmetric
extension of the standard model (MSSM). This symmetry appears anomalous, but the
anomaly is cancelled by the (discrete) Green--Schwarz (GS) 
mechanism~\cite{Green:1984sg} in such a way that it does not spoil gauge 
coupling unification (see e.g.\ \cite{Chen:2012tia} for a discussion). More 
precisely, if one extends the MSSM by a symmetry (continuous or discrete)  that
solves the $\mu$ problem and demands (i)~anomaly freedom (while allowing  GS
anomaly cancellation), (ii)~that the usual Yukawa couplings and the Weinberg 
operator are allowed, (iii) consistency with SO(10) grand unification and  (iv)
precision gauge coupling unification, then this $\Z4^R$ is the unique  solution
\cite{Lee:2010gv} (see also \cite{Chen:2014gua} for an alternative  proof). By
relaxing (iii) to consistency with SU(5), one obtains four  additional
symmetries \cite{Lee:2011dya}. Further, $\Z4^R$ can be thought of as  a discrete
remnant of the Lorentz symmetry of compact extra dimensions,  i.e.\ it has a
simple  geometric interpretation, and can arise in explicit  string--derived
models with the precise MSSM matter content \cite{Kappl:2010yu}. The charge
assignment is very simple: MSSM matter superfields have $\Z4^R$ charge 1 while
the Higgs superfields have 0, and the superpotential $\mathscr{W}$  carries $R$
charge 2. 

However, if one attempts to construct singlet extensions of the $\Z4^R$ MSSM, 
one faces the problem that the presence of a superpotential coupling of the 
singlet $\singlet$ to the Higgs bilinear $\Hu\Hd$ implies that also a linear 
term in the singlet is allowed by all symmetries.
In more detail, since the Higgs bilinear has $\Z4^R$ charge 0, the singlet 
\singlet\ needs to carry charge 2 in order to match the $\Z4^R$ charge 2 of
the superpotential. Then the desired term $\mathscr{W} \subset 
\singlet\,\Hu\Hd$ is allowed. However, in this case one might expect to have 
a problematic, unsuppressed linear term in \singlet\ in the (effective) 
superpotential,
\begin{equation}\label{eq:LinearTerm}
 \mathscr{W}_\mathrm{eff}~\supset~\Lambda^2\,\singlet
\end{equation}
with $\Lambda$ of the order of the fundamental scale. In order to forbid this 
linear term, one may try to add a new symmetry. It is quite straightforward to
see that an ordinary symmetry cannot forbid this linear term and be, at the same
time, consistent with the criteria (i)--(iv) above: in order to forbid the
linear term, the singlet \singlet\ needs to carry a non--trivial charge under
the new symmetry. But, as we want the term $\singlet\,\Hu\Hd$, this  implies
that also $\Hu\Hd$ carries a non--trivial charge. Consequently, the new 
symmetry would yield a solution to the $\mu$ problem. However, this is 
not possible: as stated above, one can prove that (under our assumptions) the 
unique solution to the $\mu$ problem is $\Z4^R$, and this symmetry does not 
forbid the linear term.

In this note, we take an alternative route and describe how one can get rid of
the linear term \eqref{eq:LinearTerm} by employing holomorphic zeros
\cite{Leurer:1992wg} associated to an additional  pseudo--anomalous $\U{1}$
gauge symmetry.

\section{Forbidding the linear term in the
\texorpdfstring{$\boldsymbol{\Z4^R}$}{R} (G)NMSSM}

\subsection{Setup}

Consider a singlet extension of the MSSM with a singlet $\singlet$ and an 
additional $\Z4^R\times\U1_\mathrm{anom}$ symmetry. $\U1_\mathrm{anom}$ is a
pseudo--anomalous \U1 symmetry, whose anomaly is cancelled by the GS mechanism.
Such \U1 factors often arise in string compactifications, and are accompanied by
a non--trivial Fayet--Iliopoulos (FI) term~\cite{Fayet:1974jb} $\xi$, which
arises at 1--loop~\cite{Fischler:1981zk}. The FI term of the $\U1_\mathrm{anom}$
is assumed to be cancelled by a non--trivial vacuum expectation value (VEV)  of
a `flavon' $\phi$, which carries negative $\U1_\mathrm{anom}$ charge and $\Z4^R$
charge 0. Without loss of generality, we can normalize $\U1_\mathrm{anom}$ such
that $\phi$ has charge $-1$ and $\xi>0$.\footnote{Of course, in true
string--derived models the situation is usually more complicated: in
approximately 500 out of a total of 11940 MSSM--like models from
\cite{Nilles:2014owa} the FI term can be cancelled with one field only. In all
other models, one would have to identify $\phi$ with an appropriate monomial of
MSSM singlet fields (see  \Apref{app:FIterm} for details).} For the sake of
definiteness, we assume that 
\begin{equation}
 \varepsilon~:=~\frac{\langle\phi\rangle}{M_\mathrm{P}} ~\sim~ 
 \sin \vartheta_\mathrm{Cabibbo} ~\sim~ 0.2\;,
\end{equation}
where the Planck scale $M_\mathrm{P}$ is identified with the `fundamental
scale'. In this case, $\U1_\mathrm{anom}$ can be used as Froggatt--Nielsen 
symmetry \cite{Froggatt:1978nt} to explain the flavor structure of quarks and 
leptons. However, this assumption is not crucial for the subsequent discussion, 
yet this is what one gets in explicit orbifold compactifications of the 
heterotic string which exhibit the exact MSSM spectrum at energies below the 
compactification scale.

Further, also the anomaly of $\Z4^R$ is assumed to be  cancelled by the GS
mechanism with the GS axion being contained in the dilaton or another
superfield, which we will denote by $\dilaton$. Since the mixed
$\U1_\mathrm{anom}-G_\mathrm{SM}^2$ and $\Z4^R-G_\mathrm{SM}^2$ anomalies are
universal, the GS mechanism does not interfere with the beautiful picture of
MSSM gauge coupling unification (see e.g.\ \cite{Chen:2012tia}).  The
`non--perturbative' term $\mathrm{e}^{-b\,\dilaton}$ carries the same $\Z4^R$
charge as the superpotential, namely 2. It might be thought of as some
non--perturbative hidden sector (see e.g.~\cite{Binetruy:1996uv}). Further,
$\mathrm{e}^{-b\,\dilaton}$ will also carry positive  $\U1_\mathrm{anom}$ charge
$s>0$ such that holomorphic zeros get lifted by `non--perturbative' terms. More
details on the charge of $\mathrm{e}^{-b\,\dilaton}$ can be found in
\Apref{app:nonpert} (see e.g.\  \cite{ArkaniHamed:1998nu,Lee:2011dya}).
In more detail, we demand that 
\begin{equation}\label{eq:Whid}
 \mathscr{W}_\mathrm{hid}~\sim~\phi^s\,\mathrm{e}^{-b\,\dilaton}
\end{equation}
be allowed, which is equivalent to the statement that $\mathrm{e}^{-b\,\dilaton}$
carries $\U1_\mathrm{anom}$ charge $s>0$.\footnote{Note that $s$ may also be
fractional even if the charges of all `fundamental ' fields are integer, for
instance if one assumes that $\mathscr{W}_\mathrm{hid}$ is given by the
Affleck--Dine--Seiberg superpotential~\cite{Affleck:1983mk}. Examples for
such terms can be found e.g.\ in \cite{Binetruy:1996uv}.} 
$\mathscr{W}_\mathrm{hid}$ may be thought of as gaugino 
condensate~\cite{Nilles:1982ik} or some other non--perturbative physics, such 
as the one discussed in~\cite{Intriligator:2006dd}, that is involved in 
spontaneous supersymmetry breaking. We discuss this in more detail in 
\Apref{app:nonpert}. Inserting the $\phi$ VEV we obtain
\begin{equation}\label{eq:Whid2}
 \mathscr{W}_\mathrm{hid}~\xrightarrow{\phi\to\Braket{\phi}}~m_{\nicefrac{3}{2}}
\end{equation}
in Planck units.\footnote{Note that \eqref{eq:Whid}  is not the `full' hidden
sector superpotential. One must, of course, make sure that $\phi$ does not
attain an $F$--term VEV, and one needs to cancel the vacuum energy. A detailed
discussion of these issues is, however, beyond the scope of the present note.}
This implies, in particular, that
\begin{equation}
 \Braket{\mathrm{e}^{-b\,\dilaton}}~\sim~\frac{m_{\nicefrac{3}{2}}}{\varepsilon^s}\;.
\end{equation}
That is, $R$ symmetry breaking is controlled by the gravitino mass, as it should
be, and due to the presence of $\U1_\mathrm{anom}$ we obtain a
Froggatt--Nielsen--like \cite{Froggatt:1978nt} modification of the terms. 
However, in contrast to the usual Froggatt--Nielsen mechanism, it yields in our
setup an enhancement rather than a suppression factor for the lifting of the 
holomorphic zeros by non--perturbative effects.

\subsection{Charges and allowed terms in the superpotential}

We summarize the $\U1_\mathrm{anom}$ and $\Z4^R$ charges in
\Tabref{tab:ChargeAssignment}.

\begin{table}[!h!]
\centerline{\begin{tabular}{|l|c|c|c|c|}
\hline
 & $\phi$ &  $\Hu\Hd$ & \singlet &  $\mathrm{e}^{-b\,\dilaton}$ \\
\hline
$\U1_\mathrm{anom}$ &
 $-1$ & $h>0$ & $n<0$ & $s>0$ \\
$\Z4^R$ & 0 & 0 & $2$ & $2$\\
\hline 
\end{tabular}}
\caption{Charge assignment.}
\label{tab:ChargeAssignment}
\end{table}

Below the $\U1_\mathrm{anom}$ breaking scale set by the $\phi$ VEV, we wish  to
have a non--trivial $\mu$ term at the non--perturbative level, i.e.\
\begin{equation}
 \mathscr{W}_\mathrm{eff}~\supset~\phi^{s+h}\, \mathrm{e}^{-b\,\dilaton}\,\Hu\Hd\;.
\end{equation}
This implies
\begin{equation}
 s+h~\ge~0 \;.
\end{equation}
We will then get effectively
\begin{equation}
 \mathscr{W}_\mathrm{eff}~\supset~M_\mathrm{P}\,\mathrm{e}^{-b\,\dilaton}\,
 \left(\frac{\Braket{\phi}}{M_\mathrm{P}}\right)^{s+h}\, \Hu\Hd
 ~=:~
 \mu\,\Hu\Hd\qquad\text{with}~
 \mu~\sim~ m_{\nicefrac{3}{2}}\,\varepsilon^{h}\;.
\end{equation}
Next, we wish to couple the singlet \singlet\ to the Higgs bilinear. We
hence demand that
\begin{equation}
 n+h~\ge~0
\end{equation}
such that
\begin{equation}
 \mathscr{W}_\mathrm{eff}~\supset~
 \left(\frac{\Braket{\phi}}{M_\mathrm{P}}\right)^{n+h}
 \singlet\,\Hu\Hd
 ~\sim~
 \varepsilon^{n+h}\,\singlet\,\Hu\Hd
 ~=:~
 \lambda\,\singlet\,\Hu\Hd\quad\text{with}~
 \lambda~\sim~ \varepsilon^{n+h}\;.
\end{equation}
Now we wish to forbid the linear term in \singlet\ at the perturbative level. 
This can be achieved with holomorphic zeros \cite{Leurer:1992wg}, 
which amounts in our setup to demanding that
\begin{equation}
 n~\stackrel{!}{<}~0
 \;.
\end{equation}
This implies, in particular, that the cubic term in \singlet\ is also forbidden.

Of course, this all works only if we make sure that $\phi$ rather than \singlet\
cancels the FI term. This might be achieved by postulating that the soft mass
squared of \singlet\ is positive while the one of $\phi$ is negative, i.e.\
\begin{equation}
 \widetilde{m}^2_\phi~<~0
 \quad\text{and}\quad
 \widetilde{m}^2_\singlet~>~0\;.
\end{equation}
A full justification of such an assumption would require to derive the setting
from some UV complete construction such as a string model. This is, however,
beyond the scope of this note.

We further obtain non--perturbative terms which are linear or quadratic in
\singlet\ if $n+2s\ge0$ or $2n+s\ge0$, respectively. Altogether we have
\begin{subequations}
\begin{align}
 n + h &~\ge~0 & &~\Leftrightarrow~\text{coupling $\lambda$ between \singlet\ 
 and $\Hu\Hd$ with $\lambda\sim\varepsilon^{n+h}$}\;,\\
 n &~<~ 0 & &~\Leftrightarrow~\text{forbid linear term in \singlet}\;,
 \label{eq:ForbidLinearTerm}
 \\
 s + h &~\ge~0& &~\Leftrightarrow~\text{$\mu$ term with
 $\mu\sim M_\mathrm{P}\,\varepsilon^{s+h}\,\mathrm{e}^{-b\,\dilaton}
 \sim\varepsilon^{h}\,m_{\nicefrac{3}{2}}$}\;,\\
 n + 2 s &~\ge~0& &~\Leftrightarrow~\text{$f^2\,\singlet$ term with $f\sim
 M_\mathrm{P}\, \varepsilon^{(n+2s)/2}\,\mathrm{e}^{-b\,\dilaton}
 \sim\varepsilon^{n/2}\,m_{\nicefrac{3}{2}}$}\;,\\ 
 2 n + s &~\ge~0& &~\Leftrightarrow~\text{$\mu_{\singlet}\,\singlet^2$ term
 with $\mu_\singlet\sim M_\mathrm{P}\,\varepsilon^{2n+s}\,\mathrm{e}^{-b\,\dilaton}
 \sim \varepsilon^{2n}\,m_{\nicefrac{3}{2}}$}\;,\label{eq:muN}\\
 3 n + 2 s &~\ge~0& &~\Leftrightarrow~\text{$\kappa\,\singlet^3$ term
 with $\kappa\sim\varepsilon^{3n+2s}\,\left(\mathrm{e}^{-b\,\dilaton}\right)^2
 \sim \varepsilon^{3n}\,\frac{m_{\nicefrac{3}{2}}^2}{M_\mathrm{P}^2}$}\;,
\end{align}
\end{subequations}
where the coefficient $\kappa$ of the cubic term is generically highly 
suppressed. Not all conditions on $\{n,h,s\}$ are independent, e.g.\ if the 
quadratic term is allowed also, since $s>0$, the linear term will be
present.

There are many possible values that satisfy all the constraints, for instance
$\{n,h,s\}=\{-1,1,2\}$, which gives us
\begin{equation}
 \lambda~\sim~\mathscr{O}(1)
 \;,\quad
 \mu~\sim~\varepsilon\,m_{\nicefrac{3}{2}}
 \;,\quad
 \mu_\singlet~\sim~\frac{1}{\varepsilon^2}
 m_{\nicefrac{3}{2}}
 \quad\text{and}\quad
 f~\sim~\frac{1}{\sqrt{\varepsilon}}\,m_{\nicefrac{3}{2}}\;.
\end{equation}
That is, the (holomorphic) $\mu$ term is roughly two orders of magnitude smaller
than $\mu_\singlet$, which might be favorable in view of the so--called `little hierarchy
problem'.

Note also that the effective superpotential
\begin{equation}\label{eq:Weff}
\mathscr{W}_\mathrm{eff}~=~
f^2\, \singlet+\mu\,\Hd \Hu  +\lambda\,   \singlet \,\Hd \Hu 
+\mu_\singlet\, \singlet^2
\end{equation}
admits two solutions to the $F$- and $D$--term equations, the first one being 
(recall that $n<0$)
\begin{subequations}
\begin{eqnarray}
 \Braket{\singlet} & = & -\frac{\mu }{\lambda } ~\sim~ -\varepsilon^{|n|}\, m_{\nicefrac{3}{2}}\;,\\
 \Braket{\Hu} & = & \Braket{\Hd}
 ~=~
  \frac{\sqrt{2 \mu  \,\mu_{\singlet}-\lambda\,f^2}}{\lambda } 
  ~\sim~ \varepsilon^{-h/2}\, m_{\nicefrac{3}{2}}\;.
\end{eqnarray}
\end{subequations}
Here one has electroweak symmetry breaking prior to supersymmetry breaking, and
the Higgs VEV may be subject to cancellations since both $\mu\,\mu_{\singlet}$
and $\lambda\,f^2$ are of the order  $\varepsilon^{2n+h}$,  e.g.\
$\varepsilon^{-1}$ in our example.
The second solution is
\begin{subequations}
\begin{eqnarray}
   \Braket{\singlet} & =& -\frac{f^2}{2\,\mu_\singlet} 
   ~\sim~ -\frac{1}{2}\,\varepsilon^{|n|}\, m_{\nicefrac{3}{2}}\;,\\
   \Braket{\Hu}&=& \Braket{\Hd}~=~ 0
\end{eqnarray}
\end{subequations}
with unbroken electroweak symmetry for unbroken supersymmetry.

\subsection{Discussion}

In summary, we find that the $\Z4^R\times\U1_\mathrm{anom}$ charge assignment 
of \Tabref{tab:ChargeAssignment} yields an effective superpotential, 
\begin{equation}
\mathscr{W}_\mathrm{eff}~=~
f^2\, \singlet+\mu\,\Hd \Hu  +\lambda\,   \singlet \,\Hd \Hu 
+\mu_\singlet\, \singlet^2
\end{equation}
with all the dimensionful parameters $\mu$, $\mu_\singlet$ and $f$ of the
order of the gravitino mass $m_{\nicefrac{3}{2}}$. This description is valid 
below the $\U1_\mathrm{anom}$ breaking scale, which is set by the flavon VEV 
$\Braket{\phi}$. In particular, the linear term in the singlet is sufficiently 
suppressed. In contrast to the original GNMSSM~\cite{Drees:1988fc}, here 
\begin{itemize}
 \item there is (essentially) no cubic term in \singlet\
 \item there is a suppressed linear term in \singlet.\footnote{Note that, unlike
 in \cite{Drees:1988fc}, we cannot shift the singlet in order to eliminate
 the linear term because the point
 $\singlet=0$ is special as it denotes the point of unbroken $\Z4^R$.}
\end{itemize} 
The scheme leads to certain predictions and expectations:
\begin{enumerate}
 \item Forbidding the linear term by holomorphic zeros implies the absence of a
 perturbative cubic term in \singlet.
 \item Further, we obtain the `little hierarchies' (recall that $n<0$)
\begin{equation}
 \mu~\sim~\varepsilon^{h+2|n|}\,\mu_\singlet
 \quad\text{and}\quad
 f~\sim~\mu/\varepsilon^{h+|n|/2}\;.
\end{equation}
\end{enumerate}

\subsection{Further applications}

Clearly, this method of avoiding a linear term in a gauge singlet may find
further applications. For instance, in model building one sometimes introduces
so--called ``driving fields'' in order to ``explain'' a certain structure of
flavon VEVs. Here, one may forbid too large tadpole terms in the same way as 
we have discussed above. 

\section{Discussion}

We have discussed how to build singlet extensions of the MSSM with $\Z4^R$
symmetry. We have shown that a potentially large linear term in the singlet can
be avoided by using holomorphic zeros. The resulting model has a $\mu$ term, a
supersymmetric mass of the order of the gravitino mass $m_{\nicefrac{3}{2}}$, as
well as a coefficient of an effective linear term in the singlet of the order
$m_{\nicefrac{3}{2}}^2$. $\mu$ is expected to be one or two orders of magnitude
smaller than $\mu_\singlet$. This might be viewed as the first step towards a solution
to the little hierarchy problem, i.e.\ explain why the electroweak scale is at
least one order of magnitude smaller than the soft supersymmetric terms. To
obtain a complete solution requires the derivation of our setting from a UV
complete model, which allows us to compute the various terms precisely. This,
however, is beyond the scope of this note.

\subsection*{Acknowledgments}

We would like to thank Mu--Chun Chen and Graham Ross for useful discussions.
M.R.\  would like to thank the UC Irvine, where part of this work was done, for
hospitality. This work was partially supported by the DFG cluster of excellence
``Origin and Structure of the Universe'' (www.universe-cluster.de) by  Deutsche
Forschungsgemeinschaft (DFG).  The authors would like to thank the Aspen Center
for Physics for  hospitality and support. This research was done in the context
of the ERC  Advanced Grant project ``FLAVOUR''~(267104).

\appendix

\section{Cancellation of the FI term}
\label{app:FIterm}

In this appendix, we discuss how the FI term gets cancelled by a single 
monomial $\mathscr{M}$. The generalization to the case of several monomials is 
straightforward. We consider a monomial of chiral superfields $\phi_i$, which 
are assumed to be standard model singlets,
\begin{equation}
\mathscr{M}~=~\prod_i \phi_i^{n_i}\;,
\end{equation}
with $n_i \in\mathbb{N}$. $\mathscr{M}$ is constructed to be gauge invariant 
with respect to all gauge symmetries except the `anomalous' $\U1_\mathrm{anom}$. 
In a supersymmetric vacuum one then has
\begin{equation}
 \frac{\left|\Braket{\phi_i}\right|}{\sqrt{n_i}}~=~v\;,
\end{equation}
where $v$ is determined from the requirement that the FI term $\xi > 0$ in the 
$D$--term potential of the anomalous $\U1_\mathrm{anom}$ gets cancelled. That
is,
\begin{equation}
 0~\stackrel{!}{=}~ D_\text{anom} ~=~ \xi + \sum\limits_i 
 Q_\mathrm{anom}^{(i)} \,|\Braket{\phi_i}|^2~=~ \xi + v^2\,\sum\limits_i 
 Q_\mathrm{anom}^{(i)}\,n_i\;,
\end{equation}
i.e.\
\begin{equation}
v~=~\sqrt{-\frac{\xi}{\displaystyle\sum\limits_i 
Q_\mathrm{anom}^{(i)} n_i}}\;.
\end{equation}
On the other hand, the `anomalous' charge of the monomial $\mathscr{M}$ is
\begin{equation}
 Q_\mathrm{anom}(\mathscr{M})~=~\sum\limits_i 
 Q_\mathrm{anom}^{(i)}\,n_i ~<~ 0\;.
\end{equation}
Hence, we obtain
\begin{equation}
 \left|\Braket{\phi_j}\right|~=~\sqrt{n_j}\,
 \sqrt{-\frac{\xi}{Q_\mathrm{anom}(\mathscr{M})}}\;.
\end{equation}
That is, if one compares the cases in which (i)~the FI term $\xi$ is cancelled
by a single field and (ii)~the FI term is cancelled by a monomial, 
there are $\sqrt{n_j}$ factors that enhance the flavon VEVs somewhat in case
(ii).

\section{Non--perturbative terms in the superpotential}
\label{app:nonpert}

In this appendix we discuss how to compute the $\U1_\mathrm{anom}$ charge of the 
non--perturbative term $\mathrm{e}^{-b\,\dilaton}$ in the case that the anomaly 
of $\U1_\mathrm{anom}$ is cancelled via the universal Green--Schwarz mechanism. 
We follow the notation of Appendix A.2 in~\cite{Lee:2011dya}.

The K\"ahler potential of the dilaton $S$ reads 
\begin{equation}
K(S,S^\dagger,V) ~=~ -\text{ln}\left(S + S^\dagger - \delta_\text{GS}\, V\right)
\end{equation}
Then, under a $\U1_\mathrm{anom}$ gauge transformation with gauge parameter 
$\Lambda(x)$, the $\U1_\mathrm{anom}$ vector field $V$ and the dilaton $S$
shift according to
\begin{subequations}
\begin{eqnarray}
V & \mapsto & V + \frac{\I}{2} \left(\Lambda(x) - \Lambda(x)^\dagger\right)\;,\\
S & \mapsto & S + \frac{\I}{2} \delta_\text{GS} \Lambda(x)\;,
\end{eqnarray}
\end{subequations}
such that $K(S,S^\dagger,V)$ is invariant. Furthermore, in order to cancel the 
cubic anomaly $A_{\U1^3_\mathrm{anom}}$, the constant $\delta_\text{GS}$ has to 
satisfy
\begin{equation}
\delta_\text{GS} ~=~ \frac{1}{2\pi^2} A_{\U1^3_\mathrm{anom}} ~=~ \frac{1}{6\pi^2} \tr Q^3_\mathrm{anom}\;,
\end{equation}
where the trace sums over the $\U1_\mathrm{anom}$ charges of all matter 
superfields. Consequently, one can define a charge $s$ for the
non--perturbative term,
\begin{equation}
\mathrm{e}^{-b\,\dilaton} ~\mapsto~ \mathrm{e}^{-\I\, s\,\Lambda(x)}\, 
\mathrm{e}^{-b\,\dilaton}\;,
\end{equation}
with $b>0$ and the charge $s$ is given by
\begin{equation}
s ~=~ Q_\mathrm{anom}\left(\mathrm{e}^{-b\,S}\right)
~=~\frac{b}{2}\delta_\mathrm{GS}~=~\frac{b}{12\pi^2} \tr Q^3_\text{anom}\;.
\end{equation}
Depending on $\tr Q^3_\text{anom}$ the charge $s$ of 
$\mathrm{e}^{-b\,\dilaton}$ can be positive or negative. On the other hand, 
in certain string--derived models, in which the Green--Schwarz
mechanism is universal, one has the relation
\begin{equation}
\label{eq:anomalyuniversality}
\tr Q_\mathrm{anom}^3 ~=~ \frac{1}{8}\tr Q_\mathrm{anom}\;,
\end{equation}
using the fact that the generator of $\U1_\mathrm{anom}$ is normalized to
$\nicefrac{1}{2}$. Then one obtains
\begin{equation}
 s~=~\frac{b}{96\pi^2}\tr Q_\mathrm{anom}\;.
\end{equation}
We have chosen $\U1_\mathrm{anom}$ such that the FI term $\xi$ is positive, 
i.e.\
\begin{equation}
\xi ~=~ \frac{g}{192\pi^2} \tr Q_\mathrm{anom} ~>~ 0\;,
\end{equation}
see Appendix \ref{app:FIterm}. Consequently, the $\U1_\mathrm{anom}$ charge of 
the non--perturbative term $\mathrm{e}^{-b\,\dilaton}$ is positive as well, i.e.
\begin{equation}
s ~=~ \frac{2b}{g} \xi ~>~ 0\;.
\end{equation}
For instance, in the case of a condensing $\SU{N_c}$ group with $N_f<N_c$
fundamental and anti--fundamental `matter' fields, $Q$ and $\widetilde{Q}$, one
has (see e.g.\ \cite[Equation~(2.7)]{Binetruy:1996uv} of the published
version)
\begin{equation}
 \mathscr{W}_\mathrm{hid}~\supset~
 (N_c-N_f)\,\frac{\Lambda^{\frac{3Nc-N_f}{N_c-N_f}}}{(\det M)^{\frac{1}{N_c-N_f}}}
 +\left(\frac{\phi}{M_\mathrm{P}}\right)^{q+\widetilde{q}}\, m_i^{\bar\jmath}\,
 M_{\bar\jmath}^i
 \;,
\end{equation}
where $\Lambda$ denotes the renormalization group invariant scale and carries
charge $Q_\mathrm{anom}(\Lambda)=N_f\frac{q+\widetilde{q}}{3N_c-N_f}$.
$q$ and $\widetilde{q}$ are the `anomalous' charges of $Q$ and
$\widetilde{Q}$, respectively. Inserting the VEV of the mesons
$M^i_{\bar\jmath}=Q^i\widetilde{Q}_{\bar\jmath}$
(see~\cite[Equation~(2.13)]{Binetruy:1996uv}), one obtains a term of the form
\eqref{eq:Whid}.

\bibliography{Orbifold}

\providecommand{\bysame}{\leavevmode\hbox to3em{\hrulefill}\thinspace}
\frenchspacing
\newcommand{\origttfamily}{}
\let\origttfamily=\ttfamily
\renewcommand{\ttfamily}{\origttfamily \hyphenchar\font=`\-}

\begin{thebibliography}{10}

\bibitem{Babu:2002tx}
K.~Babu, I.~Gogoladze, and K.~Wang, Nucl. Phys. \textbf{B660} (2003), 322,
  \texttt{arXiv:hep-ph/0212245} [hep-ph].
%%CITATION = HEP-PH/0212245;%%

\bibitem{Lee:2010gv}
H.~M. Lee, S.~Raby, M.~Ratz, G.~G. Ross, R.~Schieren, et~al., Phys. Lett.
  \textbf{B694} (2011), 491, \texttt{arXiv:1009.0905} [hep-ph].
%%CITATION = ARXIV:1009.0905;%%

\bibitem{Green:1984sg}
M.~B. Green and J.~H. Schwarz, Phys. Lett. \textbf{B149} (1984), 117.
%%CITATION = PHLTA,B149,117;%%

\bibitem{Chen:2012tia}
M.-C. Chen, M.~Fallbacher, and M.~Ratz, Mod.Phys. Lett. \textbf{A27} (2012),
  1230044, \texttt{arXiv:1211.6247} [hep-ph].
%%CITATION = ARXIV:1211.6247;%%

\bibitem{Chen:2014gua}
M.-C. Chen, M.~Ratz, and V.~Takhistov, Nucl.Phys. \textbf{B891} (2014), 322,
  \texttt{arXiv:1410.3474} [hep-ph].
%%CITATION = ARXIV:1410.3474;%%

\bibitem{Lee:2011dya}
H.~M. Lee, S.~Raby, M.~Ratz, G.~G. Ross, R.~Schieren, et~al., Nucl. Phys.
  \textbf{B850} (2011), 1, \texttt{arXiv:1102.3595} [hep-ph].
%%CITATION = ARXIV:1102.3595;%%

\bibitem{Kappl:2010yu}
R.~Kappl, B.~Petersen, S.~Raby, M.~Ratz, R.~Schieren, and P.~K. Vaudrevange,
  Nucl. Phys. \textbf{B847} (2011), 325, \texttt{arXiv:1012.4574} [hep-th].

\bibitem{Leurer:1992wg}
M.~Leurer, Y.~Nir, and N.~Seiberg, Nucl. Phys. \textbf{B398} (1993), 319,
  \texttt{arXiv:hep-ph/9212278} [hep-ph].
%%CITATION = HEP-PH/9212278;%%

\bibitem{Fayet:1974jb}
P.~Fayet and J.~Iliopoulos, Phys. Lett. \textbf{B51} (1974), 461.
%%CITATION = PHLTA,B51,461;%%

\bibitem{Fischler:1981zk}
W.~Fischler, H.~P. Nilles, J.~Polchinski, S.~Raby, and L.~Susskind, Phys. Rev.
  Lett. \textbf{47} (1981), 757.
%%CITATION = PRLTA,47,757;%%

\bibitem{Nilles:2014owa}
H.~P. Nilles and P.~K.~S. Vaudrevange, Advanced Series on Directions in High
  Energy Physics: Perspectives on String Phenomenology \textbf{Volume 22}
  (2015), \texttt{arXiv:1403.1597} [hep-th].
%%CITATION = ARXIV:1403.1597;%%

\bibitem{Froggatt:1978nt}
C.~Froggatt and H.~B. Nielsen, Nucl. Phys. \textbf{B147} (1979), 277.
%%CITATION = NUPHA,B147,277;%%

\bibitem{Binetruy:1996uv}
P.~Binetruy and E.~Dudas, Phys. Lett. \textbf{B389} (1996), 503,
  \texttt{hep-th/9607172}.
%%CITATION = HEP-TH/9607172;%%

\bibitem{ArkaniHamed:1998nu}
N.~Arkani-Hamed, M.~Dine, and S.~P. Martin, Phys. Lett. \textbf{B431} (1998),
  329, \texttt{hep-ph/9803432}.
%%CITATION = HEP-PH/9803432;%%

\bibitem{Affleck:1983mk}
I.~Affleck, M.~Dine, and N.~Seiberg, Nucl. Phys. \textbf{B241} (1984), 493.
%%CITATION = NUPHA,B241,493;%%

\bibitem{Nilles:1982ik}
H.~P. Nilles, Phys. Lett. \textbf{B115} (1982), 193.
%%CITATION = PHLTA,B115,193;%%

\bibitem{Intriligator:2006dd}
K.~Intriligator, N.~Seiberg, and D.~Shih, JHEP \textbf{04} (2006), 021,
  \texttt{hep-th/0602239}.
%%CITATION = HEP-TH/0602239;%%

\bibitem{Drees:1988fc}
M.~Drees, Int. J. Mod. Phys. \textbf{A4} (1989), 3635.
%%CITATION = IMPAE,A4,3635;%%

\end{thebibliography}
\bibliographystyle{NewArXiv} %OurBibTeX
\end{document}